\documentstyle[aps,prl,doublespace,preprint]{revtex}
\newcommand{\bb}{\begin{eqnarray}}
\newcommand{\ee}{\end{eqnarray}}
\newcommand{\p}{\partial}
\newcommand{\eps}{\epsilon}
\newcommand{\e}{\epsilon_h^*}
\begin{document}
\draft

\title{Statistical Mechanics of Membrane Protein Conformation:\\
A Homopolymer Model}
\author{Pyeong Jun Park and W. Sung}
\address{Department of Physics and Polymer Research Institute, 
Pohang University of Science and Technology\\
Pohang, 790-784, Korea}

\maketitle
\begin{abstract}
The conformation and the phase diagram of a membrane protein
are investigated via grand canonical ensemble approach
using a homopolymer model.
We discuss the nature and pathway of 
$\alpha$-helix integration into the membrane
that results depending upon membrane
permeability and polymer adsorptivity.
For a membrane with a permeability larger than a critical value,
the integration becomes the second order transition that occurs 
at the same temperature as that of the adsorption transition.
For a nonadsorbing membrane, the integration is of the first order
due to the aggregation of $\alpha$-helices.\\
\end{abstract}


\newpage

Membrane proteins regulate signal transduction and
ionic or macromolecular transport across biomembranes.
Because of their unique roles in biological functions,
their conformations and the folding pathways are
important issues in biological physics no less than the corresponding
aspects of globular protein folding.
Folding of integral membrane protein carries different
characteristics compared to globular protein folding due to the hydrophobic
environment of phospholipid membrane\cite{Popot}.
In watery solvent, the
outer surface of globular protein is usually covered with hydrophilic segments,
while the inner space is filled with hydrophobic segments to minimize
the protein-solvent interaction energy.
In contrast, membrane proteins
have hydrophobic outer regions inside the membrane
to minimize the protein-lipid interaction energy\cite{Branden}.

The three dimensional structures of a great variety of globular proteins
are experimentally known. Yet  
the structures of only a few membrane proteins 
are resolved, because 
the proteins embedded in hydrophobic membrane are difficult to
handle experimentally\cite{Branden}.
Since the structural determination of Bacteriorhodopsin\cite{experiment},
the idea has been widely accepted 
that the membrane proteins are predominantly made up of
$\alpha$-helices induced by hydrogen bonding\cite{idea}.
Unlike the globular proteins,
the membrane proteins can adopt only a few basic structures such as 
$\alpha$-helix, 
allowing more tractable theoretical approaches for membrane proteins.

While a number of theoretical studies have been done separately
on globular protein folding\cite{Folding} and polymer adsorption 
on membranes\cite{Adsorption},
there are few efforts devoted to the folding of membrane proteins
involving the surface adsorption\cite{Baumgartner}.
In this Letter, we address this problem using the statistical mechanics
via grand canonical ensemble approach.
To extract the salient features of the conformations and their transitions
from the intractable complexity characteristic of the real proteins,
we introduce a simple but tenable model: a long homopolymer
which undertakes a random walk outside the membrane regarded as planar,
and can interact with it via contact binding on its surface
and penetration into its interior(Fig.~1), as will be detailed.
Motivated by the fact that hydrogen bonding is very stable in the hydrophobic
environments\cite{Lemmon}, we assume that $\alpha$-helix structure is formed 
if and only if the segments are placed within the membrane. 
Here we neglect other secondary structures such as 
$\beta$-sheets for simplicity.
Another important observation to incorporate is that 
the $\alpha$-helices preferentially aggregate
to form a thermodynamically stable structure, called $\alpha$-helix oligomer, 
which is dominant over the dispersed $\alpha$-helices\cite{Wang}. 

In our model, an $\alpha$-helix column has fixed number of hydrogen bonds $n$
(implicitly representative of membrane thickness), with the statistical weight
$W_H \sim \sigma_h^{n} \exp(-\beta (n\eps_h+\eps_a))$,
where $\beta=1/k_BT$, 
$\sigma_h\equiv\exp(\Delta s_M/k_B)<1$.
The $\eps_h<0$ and
$\Delta s_M<0$ 
are the energy and
the entropy change associated with hydrogen bonding,
and $\eps_a<0$ is the aggregation energy per helix column.
On the membrane surface, polymer segments are allowed to be
adsorbed with the statistical weight for $k$ segments given by
$W_S \sim \sigma^k \exp(- \beta k\eps_s)$,
where $\eps_s<0$ is the segmental attraction energy, and 
$\sigma\equiv\exp(\Delta s/k_B) <1$ with $\Delta s$ the segmental
entropy change by adsorption.
Due to chain connectivity, 
the domains other than the membrane-bound oligomer and surface-adsorbed trains
consist of two end tails, 
loops starting or ending at surface trains, and  
loops that connect two helix columns(return to
the starting point as the radius of an $\alpha$-helix will be neglected
in this work)(Fig.~1).
A tail of $k$ segments has the statistical weight of random walk
in half space, which departs from the membrane surface and ends up anywhere
in the half space\cite{Sung},
$W_T \sim q^k/k^{1/2}$,
where $q$ is the segmental partition function.
Loops have the statistical weights 
of random walks ending up
anywhere on the surface, $W_L \sim q^k/k^{3/2}$, 
and of those returning to the starting point, 
$W_R \sim q^k/k^{5/2}$\cite{DiMarzio}.

We construct grand partition functions
of five different kinds of domains as follows.
The bound chains as a helix column and a surface-adsorbed train 
have respectively the partition functions,
\bb
Q_H(z_M) &=& I_h z_M^{n} \sigma_h^{n} \exp(-\beta (n\eps_h+\eps_a)),\\
Q_S(z) &=& I_s \sum_{k=1}^{\infty} z^k\sigma^k\exp(-\beta k\eps_s) \nonumber\\
       &=& I_s z\sigma\exp(-\beta\eps_s)/(1-z\sigma\exp(-\beta\eps_s)),
\ee
where $I_h$, $I_s$ are the nucleation or initiation parameters for
an $\alpha$-helix and a surface-adsorbed train respectively\cite{Zimm}.
The $z$ and $z_M$ are the segmental fugacities outside and inside membrane 
respectively, which defines the chemical potential difference 
$\Delta\mu\equiv\mu_M-\mu=\beta^{-1}\log(z_M/z)$, 
a measure of membrane permeability determined by environmental effects
such as membrane hydrophobicity.
With $q=1$, neglecting the irrelevant bulk contribution,
the grand partition functions 
of a tail, a loop, and a returning loop are given by
\bb
Q_T(z) &=& A_T\sum_{k=1}^{\infty} z^k/k^{1/2} \equiv A_T g_{1/2}(z)\\
Q_L(z) &=& A_L\sum_{k=1}^{\infty} z^k/k^{3/2} \equiv A_L g_{3/2}(z)\\
Q_R(z) &=& A_R\sum_{k=1}^{\infty} z^k/k^{5/2} \equiv A_R g_{5/2}(z),
\ee
where $A_T$, $A_L$, $A_R$ are constants of order of unity, 
and $g_{m}(z)$ is the polylogarithmic function of order $m$.

The total grand partition function of the membrane-bound polymer
can now be calculated considering every possible conformation
made of all the domains.
To this end, consider the transfer matrices defined as
\bb
{\bf X} = \left[ \begin{array}{cc}
			Q_S &  0  \\
			 0  & Q_H \end{array} \right],
{\bf Y} = \left[ \begin{array}{cc}
			Q_L & Q_L \\
			Q_L & Q_R \end{array} \right], \nonumber
\ee
which represent respectively the two membrane-bound domains(adsorbed-train 
and $\alpha$-helix) 
and two types of loops($L$ and $R$) joining them.
Introducing the matrix
${\bf B} = {\bf X} \sum_{p=0}^{\infty} \left( {\bf Y} {\bf X}\right)^p$, 
whose elements properly incorporate the interconnected arrays of
all the domains between two tails
(for example, the element $B_{12}$ represents the
connected arrays starting from
train and ending with helix), 
we finally get the total partition function ${\cal Q}$
and its diagrammatic representation(Fig.~2),
\bb
{\cal Q} &=&  Q_T^2(B_{11}+B_{12}+B_{21}+B_{22}) \nonumber\\
&=& \left[Q_T^2( Q_S +Q_H)\right] \nonumber \\
&+& \left[Q_T^2( Q_S^2Q_L+Q_SQ_LQ_H +Q_HQ_LQ_S+Q_H^2Q_R)\right]\nonumber\\
&+& \mbox{terms with } p\geq 2. 
\ee
Here we regarded the two ends of the polymer to be distinguishable.
We will consider the thermodynamic limit, written as $\left<N\right> 
\rightarrow \infty$, in order to define phase transitions that result.
Then the partition function is reduced to
\bb
{\cal Q} \sim \left\{ \begin{array}{ll}
		Q_T^2, & T>T_c \\
		(1-\lambda)^{-1}, & T< T_c,
		\end{array}
	\right.
\ee
where $\lambda$ is the largest eigenvalue of ${\bf Y}{\bf X}$, 
given by
\bb
\lambda = \frac{1}{2} (Q_SQ_L+Q_HQ_R)
	 +\frac{1}{2}	 \left[ (Q_SQ_L-Q_HQ_R)^2-4Q_SQ_HQ_L^2 \right]^{1/2} ,
\ee
and $T_c$ is a critical temperature determined from $\lambda(z=1,T=T_c)=1$.

For $T>T_c$, tail is the only 
allowed conformation, indicating that
the polymer tends to be desorbed to the outer region of the membrane.
For $T<T_c$, the surface trains and/or helices
with loops and returning loops in between
becomes dominant, which indicates the stability of membrane-bound phase.
The segmental fraction of each domain can be defined as
$f_i \equiv \left<N_i\right>/\left<N\right> 
= \left<N\right>^{-1}(\p\log {\cal Q}/\p \log Q_i)(\p\log Q_i/\p \log z)$,
where $i=S,R,L$ represent three different types of domains considered(Fig.~1),
and $f_H \equiv \left<N_H\right>/\left<N\right>
= \left<N\right>^{-1}(\p\log {\cal Q}/\p \log Q_H)(\p\log Q_H/\p \log z_M)$,
for helix domain.
Depicted in Fig.~3 are the segmental fractions
vs. temperature, where 
all the energy parameters, scaled in units of $|\eps_s|$, 
are taken to be the same order of magnitude,
and the entropies $\Delta s$ and $\Delta s_M$
are taken to be order of $k_B$\cite{Ross}. 
As shown in Fig.~3(a), the desorption-adsorption transition, which is of the
second order as is known, takes place at $T=T_c$,
where the order parameter, the surface-adsorbed segmental fraction($f_S$), 
increases from zero continuously as $T$ is lowered from $T_c$.
Further lowering of the temperature drives the polymer 
integration in a form of $\alpha$-helix aggregate at $T=T_h$, 
the helix inclusion temperature, which is defined by the local maximum
of specific heat.
The specific heat curve, shown in Fig.~4,
clearly indicates that structural changes occur at both $T_c$ and $T_h$.
The specific heat diverges at $T=T_c$, 
and has a local maximum at $T=T_h$ indicating the helix formation
that is a crossover\cite{crossover}.

As the value of $\e \equiv \eps_h-\Delta\mu$, the energy of
$\alpha$-helix inclusion per segment, is lowered, or membrane permeability
is increased,
$T_h$ approaches $T_c$,
so that for the values of $\e$ smaller than a critical value
(about $-2.6$ using the
parameters employed in Fig.~3), 
the helix inclusion is promoted to the second order transition with
$T_h$ and $T_c$.
Figure~3(b) depicts the segmental fractions for this case($\e=-3$).
It is shown that, in contrast to Fig.~3(a), helix formation dominates 
over adsorption below the common transition temperature.
The phase diagram in Fig.~5 summarizes 
the foregoing discussions concerning the conformational
phases and their transitions for wide range of $\e$ and temperature.

The two temperatures appearing in our model, the desorption-adsorption
temperature $T_c$ and the helix inclusion temperature $T_h$,
govern the pathway of our model-homopolymer integration in membrane.
They are respectively similar to coil-to-globule transition temperature
$T_\theta$ and folding transition temperature $T_f$ in globular protein
folding; like the globular phase, the adsorbed phase is indeed an 
intermediate state approaching the native folded structure\cite{Jacobs}.
Recently Klimov and Thirumalai showed an evidence that globular 
protein folding time is scaled as 
$\tau \sim \exp\left[J|T_\theta-T_f|/T_\theta\right]$,
where $J$ is a model-dependent constant\cite{Thirumalai}.
Even without considering the analogy, 
the Fig.~5 suggests that a transition into 
rapid $\alpha$-helix integration in membrane can be attained 
for {\em permeable and adsorbing} membranes, 
with $\e$ lower than a critical value where $T_c=T_h$.
A detailed analysis of the free energy landscape and barrier crossing dynamics
should confirm this highly plausible suggestion.

We now consider the situation of 
{\em nonadsorbing membrane} where $Q_S$, the partition function of 
surface-adsorbed domain vanishes (for example, $I_s$, the adsorption
initiation parameter is zero). We then find the eigenvalue 
$\lambda=Q_HQ_R$, signifying that the partition function incorporates
the conformations generated from recurrence of a helix and a returning loop
in series. In this case, the polymer inclusion
forming the $\alpha$-helix oligomer is found to be the first order transition.
At the transition temperature $T_h$, which is again determined by
$\lambda(z=1,T_h)=1$, the order parameter, the fraction of segments in
the $\alpha$-helix oligomer($f_H$), changes discontinuously from zero to
\bb
\Delta f_H=n/\left[n+g_{3/2}(1)/g_{5/2}(1)\right],
\ee
which incurs the segmental latent heat of dissociating 
the membrane-bound oligomer, $l_H = |\e+\eps_a/n|\Delta f_H$.
The reason why the transition should be discontinuous
in the absence of adsorption is argued as follows.
The inclusion accompanies the aggregation of helices, which 
restricts the loops to be closed between helix columns.
Presence of long returning loops is suppressed
entropically, and, also energetically in favor of the aggregate,
which tends to be of significant fraction
at the transition from the desorbed phase.
This is in sharp contrast to the two second order transitions that can be 
obtained from our theory, the desorption-adsorption 
transition without helix inclusion($\lambda=Q_SQ_L$)
and the inclusion transition in forms of dispersed helices 
without aggregation and adsorption($\lambda=Q_HQ_L$ with $\eps_a=0$),
where the loops can be transformed continuously into the adsorbed segments
and helix columns respectively.
Similar discontinuous transition was reported in the 
helix-coil transition of double strand DNA\cite{DiMarzio,DNA}.
For the biological processes without appreciable changes in temperature
and volume, certain mechanism of latent heat 
involving enzymatic activity
could be essential to facilitate the first order transition\cite{DiMarzio}.
Furthermore, the conclusions of this and foregoing paragraphs, if properly extended to an
asymmetric membrane where one side is adsorbing and the other is not\cite{Park},
imply that the rapid protein integration can be promoted only through the
adsorbing side. It would be important to confirm this possibility
as well as our results for symmetric membranes by experiments and/or
simulations.

In summary, we studied various membrane-protein conformations
and different pathways to the native structure of an $\alpha$-helix aggregate
as a function  of temperature and membrane permeability.
Two significant conclusions obtained are: 
1) The polymer inclusion pathway is determined by
membrane permeability above a critical value of which the adsorption
and the helix inclusion converge
and 2) The nature of inclusion transition is determined by
the availability of the polymer adsorption on membrane surface,
due to the chain connectivity constraint.
Although the important sequential heterogeneity and finite length effect
involving chain stiffness of real proteins are neglected,
our model gives some nonspecific features 
of membrane-protein conformation,
in particular, the roles of membrane hydrophobicity 
and segmental interaction with the membrane surface.

We acknowledge the support from KOSEF(961-0202-007-2), BSRI(97-2438), 
and POSTECH special fund program.



\begin{figure}[b]
\caption{Schematic figure of a membrane protein. Five different 
domains are indicated.}
\end{figure}

\begin{figure}[b]
\caption{Diagrammatic representation of the partition function.} 
\end{figure}

\begin{figure}
\caption{Segmental fractions of a membrane-bound polymer vs. temperature.
The parameter values throughout this paper 
are selected as[13] $\sigma = \exp(-1)$, $\sigma_h = \exp(-2)$, 
$A_L = 1$, $A_R=0.5$, $I_h = 0.01$, $I_s = 0.1$, $n=12$, $\eps_a=-3$, 
where energies and temperature are in units of $|\eps_s|$.
(a) $\e \equiv \eps_h-\Delta\mu = -2$. 
Desorption-adsorption transition takes place at $T=T_c\simeq 1.30$ 
continuously, while the helix inclusion occurs 
at $T<T_h\simeq 0.82$.
(b) $\e = -3$(more permeable membrane).
Adsorption and helix inclusion occurs simultaneously 
at $T=T_c=T_h\simeq 1.46$,
but helix structure becomes dominant over adsorbed state
at lower temperatures.}
\end{figure}

\begin{figure}
\caption{Specific heat (in units of $k_B$) versus temperature. 
Parameters are the same as in Fig.~3(a).}
\end{figure}

\begin{figure}
\caption{Phase diagram of a membrane-bound 
polymer. Solid line indicates the second order
transition, while dotted line the crossover between adsorption
and $\alpha$-helix inclusion.
Parameters are the same as in Fig.~3.}
\end{figure}

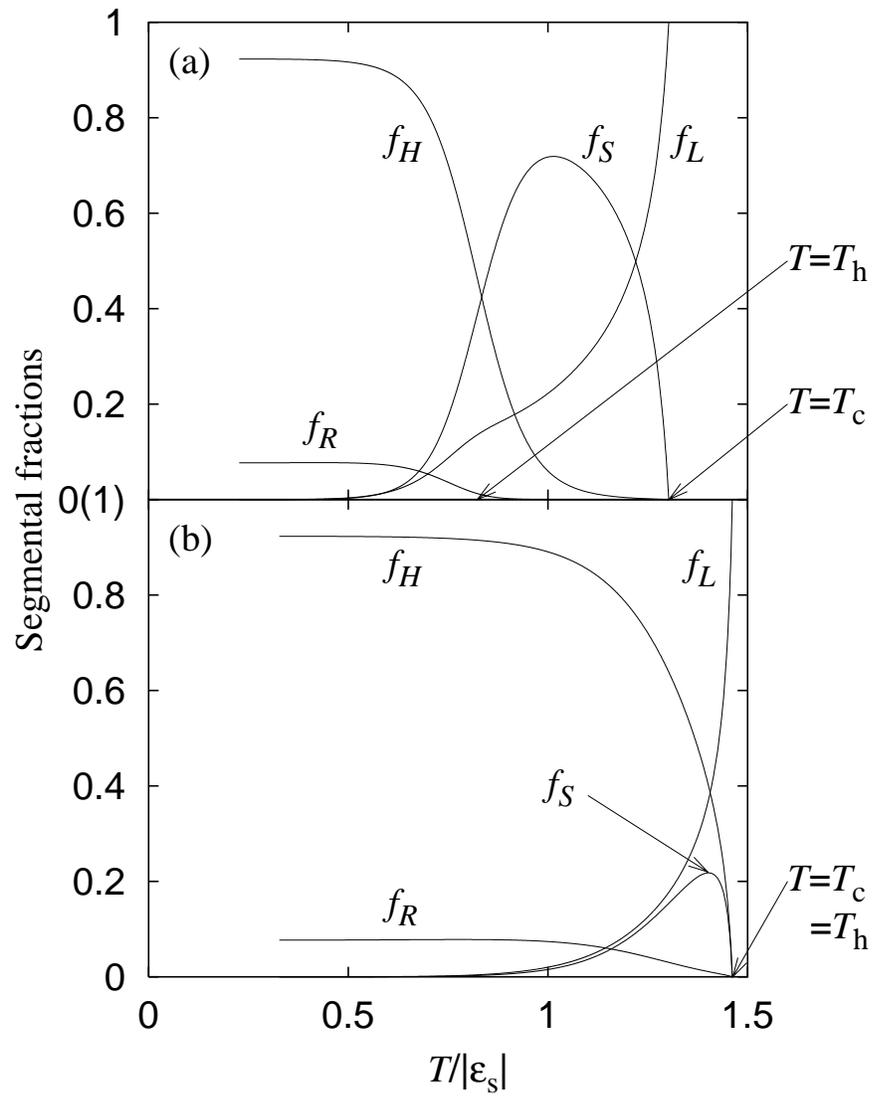
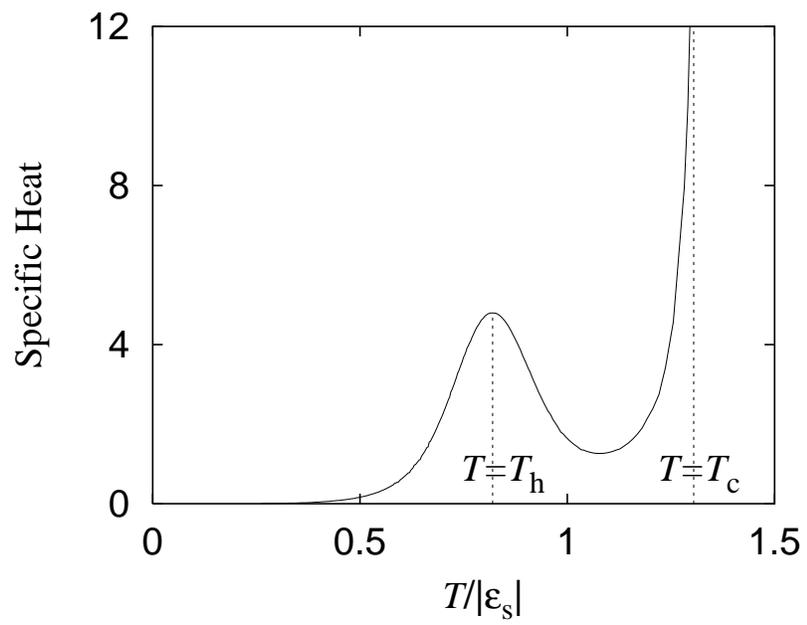
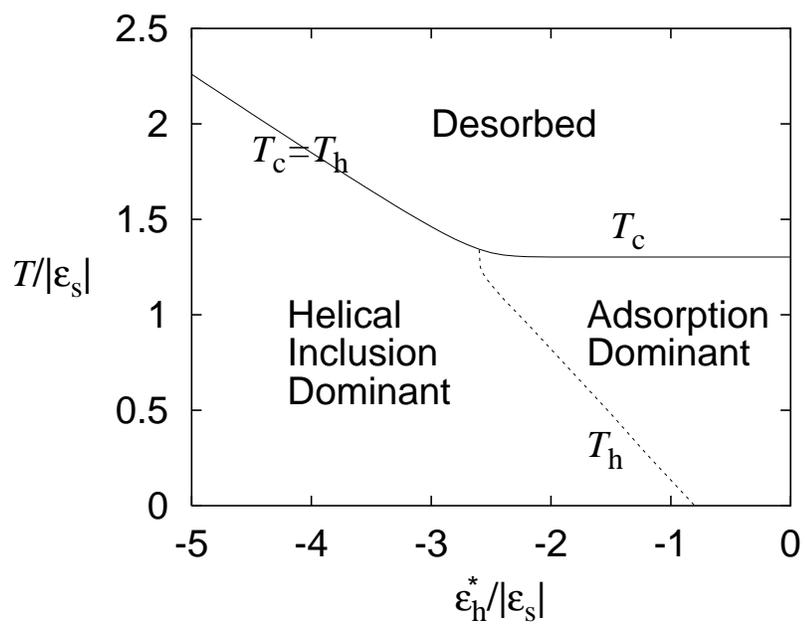
\end{document}